# Numerical Simulation of Gas Storage Caverns in Qom Region

Mostafa SHARIFZADEH, Ali MORADI GHASR

*Department of Mining, Metallurgical and Petroleum Engineering, Amirkabir Unversity of Tehran, Tehran, Iran*
*424 Hafez Avenue, P.O.Box 15875-4413, Tehran, Iran*

**Absrtact**

The rock mechanical design of gas storage cavern in salt requires the analysis of the stability and the usability of the cavern over the planned operating time period. The design includes the build up of a rock mass model and a numerical model taking into account the geological situation, load condition, geometrical condition, and material parameters. In this paper multiple caverns in salt formation with geological and geomechanical situation in Qom (central part of Iran) was investigated a using creep model. Minimum safe center to center distances (CTCD) of multiple horizontal caverns also were studied. CTCD of caverns interact at less than two times of cavern diameter. With increasing the CTCD to 2.5 times cavern diameters, diminish most interaction.

## 1. Introduction

Natural gas transportation facilities have usually limited the ability of adjusting unpredictable consumption and supply changes in a timely manner. To quickly respond to changes in natural gas demands and to ensure uninterrupted gas supplies, suitable storage facilities are often a necessity. Gas storage in underground spaces has taken great attention as a basic policy in controlling consumption markets, along with increasing growth for gas consumption as a fuel, especially in cold seasons of year. The other advantages of underground storages are the small amount of land required, and security against external influences.

Gas is stored by various means to even out pipeline loads and to take advantage of fluctuations in gas prices. Methods of underground natural gas storage include oil and gas fields, aquifers and salt caverns. Since storage in salt caverns behaves like pressurized containers, salt cavern storage has the advantage of potentially high deliverability flow rates. Furthermore, man-made salt caverns can be sized for individual needs by solution mining. Salt caverns are typically located 450 to 2500 m below surface and are kept at operating pressures of about 15 to 25 MPa.

Iran has great potential for consumption liquid gas which leads to save in crude oil production for domestic usage and will help effectively decreasing air pollution. Tehran is appropriate for gas storage among the other cities of Iran, so, salt features such as salt beds and salt domes nearby Tehran are the best candidates for investigation of locating the storing site.

Kuh-e-Namak region is formed of salt mountain area where located in north western of Qom, 100 kilometers apart from Tehran and throughout gas pipes line cut across its eastern flank, so, Kuh-e-Namak is one of the best alternative for storing site.

Gas storage simulations are investigated by Michael et al. (2) they used cylindrical shape for storage. In practice gas storage caverns are capsule shape, therefore numerical simulation with capsule shape give better result than another shape.

## 2. Geology

Bedded salt formation in all areas, however, are layered and interspersed with non-salt sedimentary materials such as anhydrite, shale, dolomite, and limestone. The salt layers themselves also often contain significant impurities. In comparison to relatively homogeneous salt domes, therefore, cavern development and operations present additional engineering challenges related to:

- The layered, heterogeneous lithology,
- Differential deformation, creep, and bedding plane slip between individual layers,
- Cavern dimensions,

There are several salt beds and salt domes, around Qom and among them, Kuh-e-Namak is the most well known and characteristic salt dome located in north west of Qom and elevated about 250m to around terrains. Salt of this region was originated from Oligocene and Miocene brines. Geological location of Kuh-e-Namak and its geological map is shown in Fig1. Several bore holes were drilled for investigating depth and thickness of salt layers in this region which their data presented in table1.

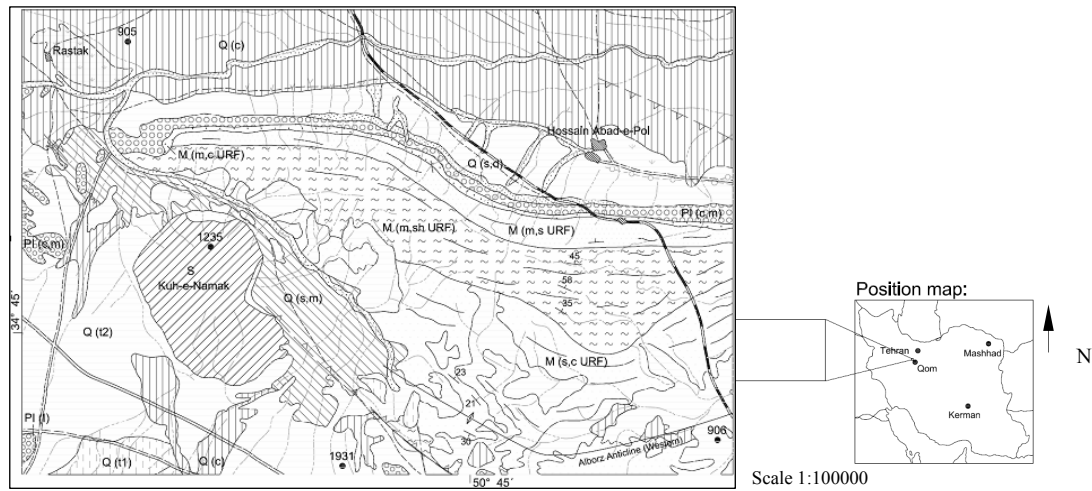

Fig1. Geological map Kuh-e-Namak in central part of Iran

Table 1. Survey summary of Kuh-e-Namak region

| Field | Holes | Upper salt (m) | Lower salt (m) | Thickness (m) |
|---|---|---|---|---|
| Alborz | Alborz 5 | 2512 | 2676.7 | 164.7 |
| | Alborz 6 | 2685.1 | 2965 | 279.9 |
| | Alborz 7 | 2654 | 2927 | 273 |
| | Alborz 8 | 2570.3 | 2784.4 | 214.1 |
| | Alborz 9 | 2435.3 | 2572.2 | 136.9 |
| | Alborz 10 | 2370 | 2549.3 | 179.3 |
| Saracheh | Saracheh 1 | 2307 | 2404.9 | 97.9 |
| | Saracheh 2 | 2350 | 2445.7 | 95.7 |
| | Saracheh 3 | 2481.1 | 2547.2 | 66.1 |
| | Saracheh 4 | 2599.3 | 2791.4 | 192.1 |
| | Saracheh 5 | 2356.4 | 2559.7 | 203.3 |
| | Saracheh 6 | 2319.8 | 2541.7 | 221.9 |
| Ave. Alborz | | 2537.78 | 2745.7 | 207.98 |
| Ave. Saracheh | | 2412.76 | 2548.43 | 146.17 |

## 3. Cavern design
### 3.1. Design fundamentals

The objective of the rock mechanical cavern design is to guarantee its sufficient stability and serviceability with respect to the planned usage.
The factors influencing the stability of an unlined underground cavity are basically divided into four groups:

- Physical effects (initial rock stresses, internal pressure, temperature)
- Rock stresses, theoretically determined with the assistance of calculation model,
- Mechanical strength of the rock that acts against the physical effects,
- Influence of geological anomalies

In cavern design, above-mentioned variables influences have to be taken into consideration during safety analysis. In particular, establishment of the partial safety coefficient, for instance, demanded by the processes underpinned by the probability theory however, it is problematic since the probability of the combined effect of different influences must be investigated together.

### 3.2. Effective parameters for gas storage cavern

The design variables to be established for the layout of a cavern are predominantly based on the final use to which the cavern is to be put, e.g. mineral extraction, storage or disposal site. The most important parameters to be taken into consideration in the design of a gas cavern are as follows:

- Depth of the cavern,
- Cavern geometry (diameter, height, roof shape),
- Distance of the caverns from one another,
- Distance of neighboring formation,
- The internal pressure conditions,

The individual caverns have a capsule shape in order to create the largest possible storage volume in the salt dome structure.
In addition to the limitation of the maximum pressure to prevent seal failures or the fracturing of the formation, the minimum allowable internal pressure during operation is also laid down, in order to avoid spaling from the walls of the cavern.

## 4. Numerical Analysis
### 4.1. Numerical simulation matrix and model description

A set of three dimensional geomechanical models to investigate cavern deformation and bedding slip for a variety of cavern configurations were developed for Kuh-e-Namak. For this purpose a computer program was written in visual basic. The gas storage-cavern program is illustrated in Fig2.
A computer graphical interface was developed to specify varying lithology layers (number, depth, thickness) and varying cavern geometry parameters (depth, height, and radius). This program has also the ability of determining the kind of injection for creative cavern and can study behavior of salt after creep.

Table 2 summarizes the initial set of geomechanical parameters obtained for this project. Numerical simulations include a baseline case and various scenarios for different salt roof beam thickness, a range of cavern depth, distance of caverns from together, and ratio of H/D (Height to Distance). Table 3 shows material properties used in parametric simulation that, this critical data selected based on geotechnical study and authors opinion.

Table2. Summary of the geomechanical models developed for Kuh-e-Namak

| Field | Main Roof | Bottom Depth | Ave. height(m) | Ave. Diameter (m) | H/D | Cavern Volume (m$^3$) | Shape |
|---|---|---|---|---|---|---|---|
| Alborz | 2537 | 2745 | 40 | 30 | $\frac{4}{3}$ | 35342 | Capsule |
|  |  |  | 50 | 30 | $\frac{4}{3}$ | 42411 | Capsule |
|  |  |  | 60 | 30 | $\frac{6}{3}$ | 49480 | Capsule |
| Saracheh | 2412 | 2548 | 40 | 30 | $\frac{4}{3}$ | 35342 | Capsule |
|  |  |  | 50 | 30 | $\frac{5}{3}$ | 42411 | Capsule |
|  |  |  | 60 | 30 | $\frac{6}{3}$ | 49480 | Capsule |

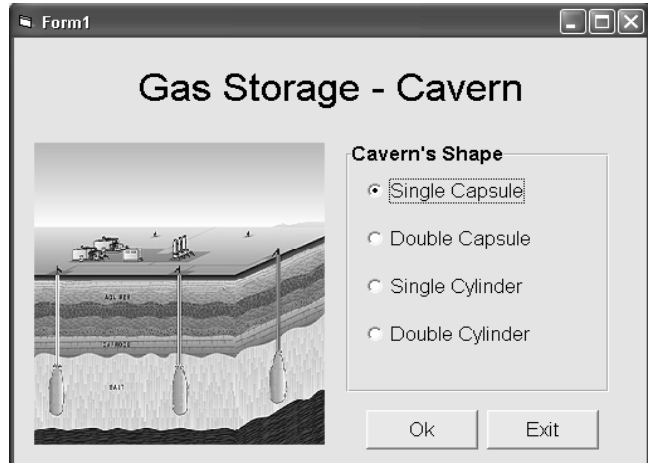
a) General interface

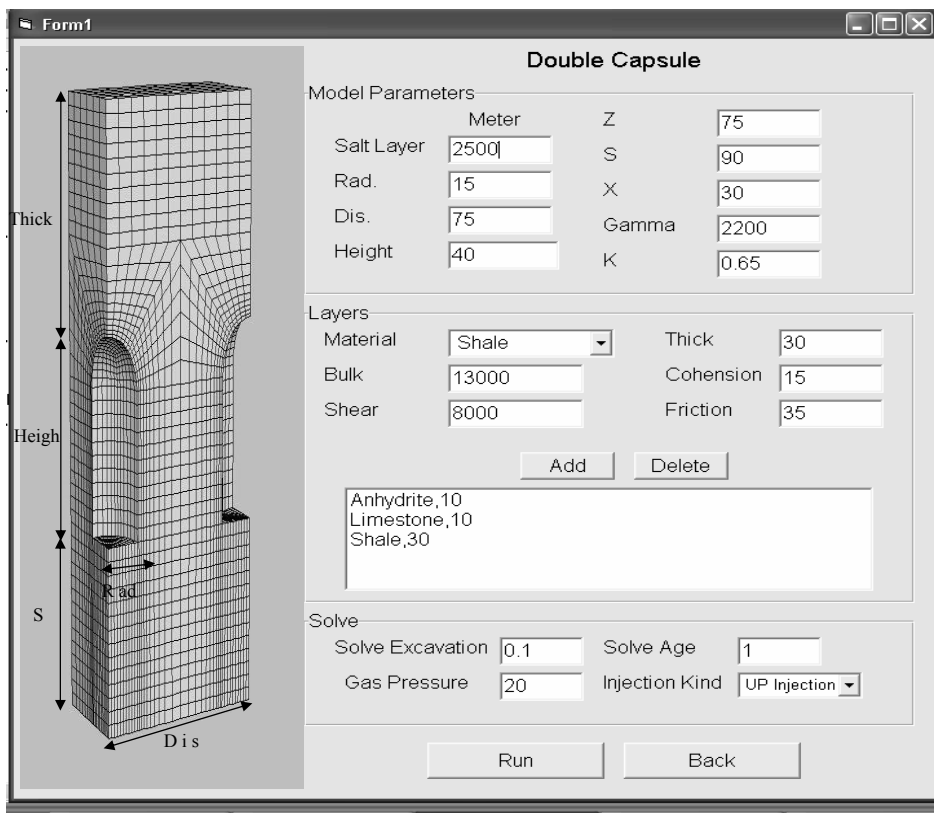
b) Input data interface

Fig2. (a), (b) Interface for analysis of Varying Cavern Configuration

Table3. Material properties used in parametric simulation

| Material | Bulk Modulus(Mpa) | Shear Modulus(Mpa) | Density (kg/m$^3$) | Tension (Mpa) | Cohension (Mpa) | Friction Angle |
|---|---|---|---|---|---|---|
| Anhydrite | 74000 | 25000 | 3000 | 7 | 20 | 35 |
| Limestone | 40000 | 25000 | 2700 | 4 | 15 | 35 |
| Shale | 13000 | 8000 | 2600 | 1 | 5 | 20 |

For each simulation a vertical stress is developed consistent with the density of overlying sediments (i.e. $\sigma_v = \int \rho g dz$ ).

Lateral displacements at the outer radius of the model are fixed, so the horizontal stresses develop consistent with the vertical load and the poisson ratio for the various lithology

layers. The general simulation process may be summarized as follows:

- Define initial geological layers and initial stress conditions,
- Cavern excavate with applying an internal cavern pressure equal to hydrostatic head of water ( about 20Mpa at depth of 2000m),
- Run code to obtain stresses to initiate creep for 1 year,

**4.2. Multiple horizontal cavern simulation**

Numerical models for a variety of multiple horizontal caverns configurations have been developed and applied to investigate cavern integrity and interaction between nearby caverns.

Table 4 summarizes the main parameters of this investigation. The geometric layout of each cavern is given by an H/D ratio, total diameter of 30m and height varying between 40, 50, and 60m. The baseline case is given by two identical horizontal caverns located at a CTCD of 75 m equal to 2.5 cavern diameter. Fig3 shows the configuration of the three dimensional model of multiple horizontal cavern baseline.

Table4. Simulation matrix for multiple horizontal caverns numerical investigations.

| No Simulation | No of caverns | Cavern height | Cavern diameter | Center distance | pressure |
|---|---|---|---|---|---|
| 1 | 2 | 40 | 30 | 75 | Hydrostatic |
|   |   | 50 | 30 |    |             |
|   |   | 60 | 30 |    |             |
| 2 | 2 | 40 | 30 | 90 | Hydrostatic |
|   |   | 50 | 30 |    |             |
|   |   | 60 | 30 |    |             |

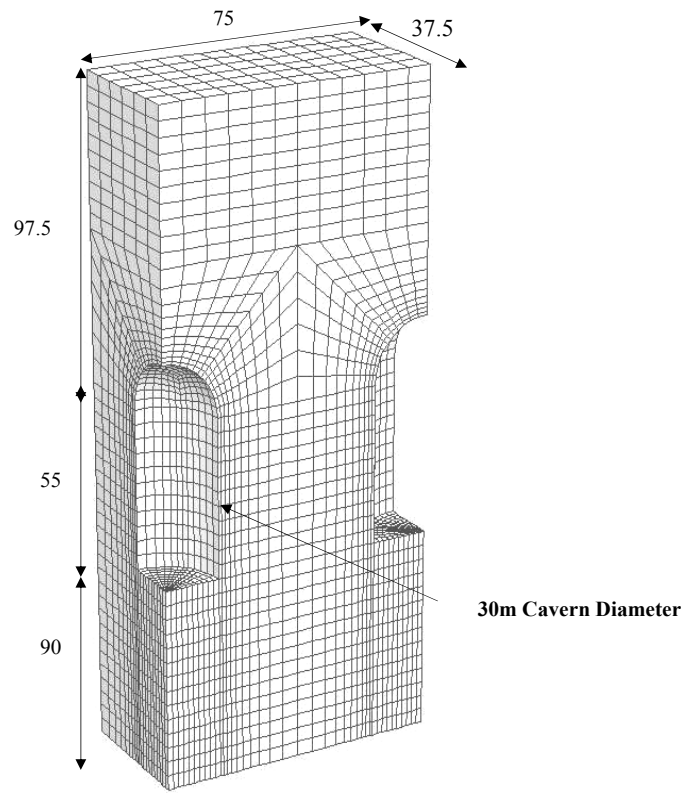

Fig3. Three dimensional multiple horizontal caverns baseline model.

## 4.2 Numerical Simulation and Analytical Results

In this paper we solve problem with numerical finite diference using Flac$^{3D}$ and analytical results. It was shown that analytical results obtained from kirsch eq. are in agreement with numerical simulation results that are given as below:

When horizontal to vertical rate is 1 (k=1), kirsch eq. convert to (1), (2), (3), and (4) equations:

$$\sigma_r = P\{1 - \frac{a^2}{r^2}\} \tag{1}$$

$$\sigma_\theta = P\{1 + \frac{a^2}{r^2}\} \tag{2}$$

$$\tau_{r\theta} = 0 \tag{3}$$

$$|P_{max} - \sigma| \leq c\% P_{max} \tag{4}$$

$$\xrightarrow{(1),(2),(3),\text{and }(4)\text{ with.}5\%.\textit{effective}} \frac{a^2}{r^2} = \frac{5}{100}$$

$$r = \sqrt{20}\ a$$

$$\Rightarrow \quad r = \sqrt{20} \times 15 \approx 67\text{m}$$

In these eq. $\sigma_r$ is radial stress, P is vertical stress, $\sigma_\theta$ is tangent stress, $\tau_{r\theta}$ is shear stress, a is radial, and r is CTCD. Therefore using kirsch eq CTCD of caverns is obtained 67m with minimum influence excavation. These results are obtained by numerical analysis, in which, diameter is 30m, and all interactions vanish.

Figs4, 5, 6 and 7 present a summary of displacement for top, bottom, and side of the cavern and contour of vertical stress for the double capsule simulation. We select the displacement magnitude as a kinematics quantity to describe and visualize caverns interaction.

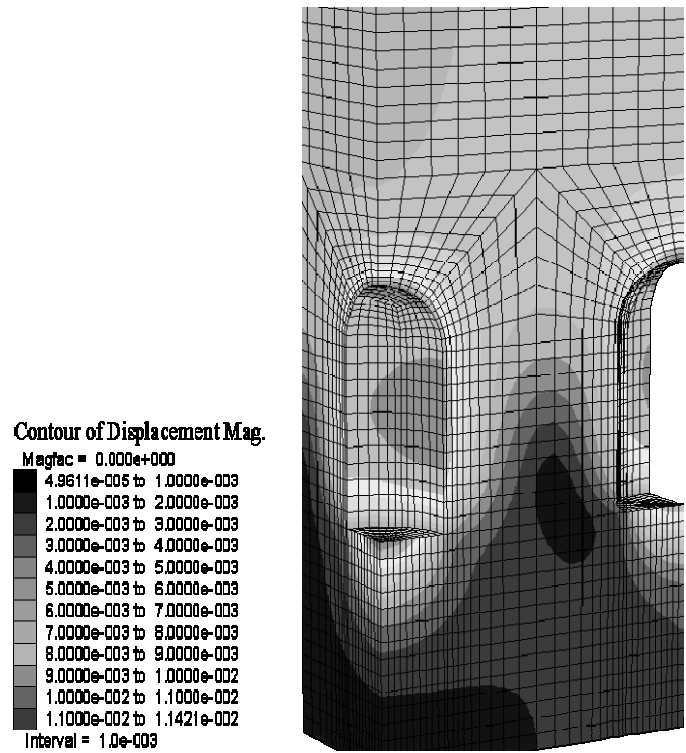

Fig4 : plot of displacement magnitude for caverns in equilibrium with cavern pressure of 20 MPa after one year. Where CTCD is 2.5 times cavern diameters.

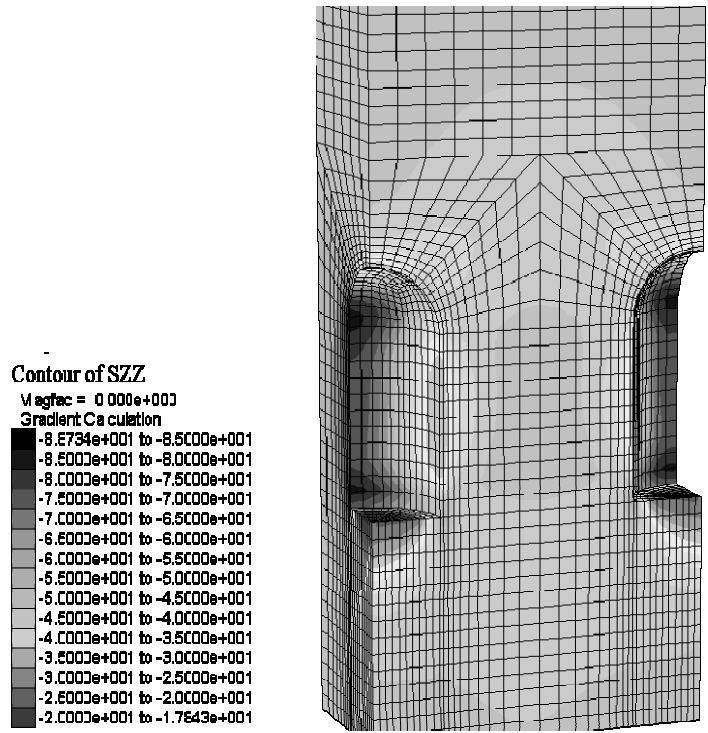

Fig5: Plot of vertical stress (Mpa) for caverns in with cavern pressure of 20 MPa after one year. Where CTCD is 2.5 time cavern diameters.

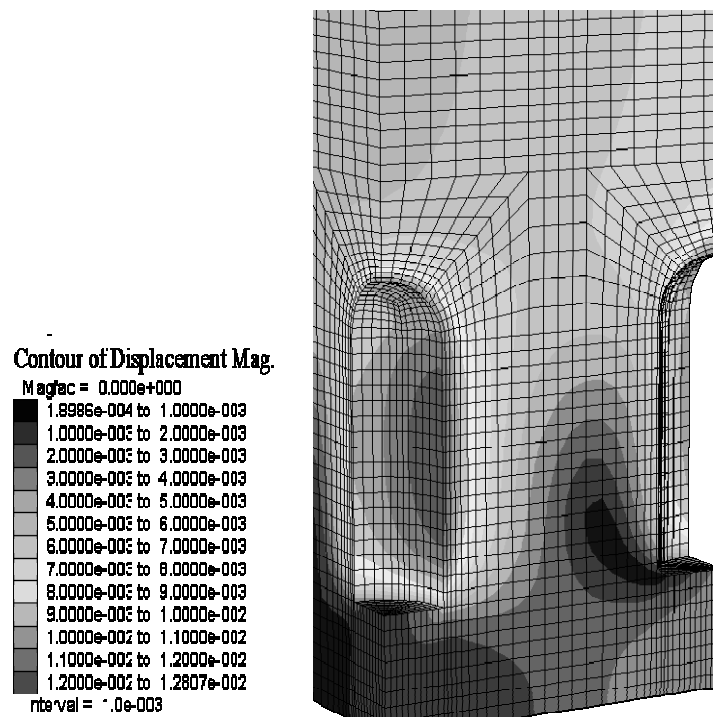

Fig6: plot of displacement with cavern pressure of 20 MPa after one year. Where CTCD is 3 time cavern diameters.

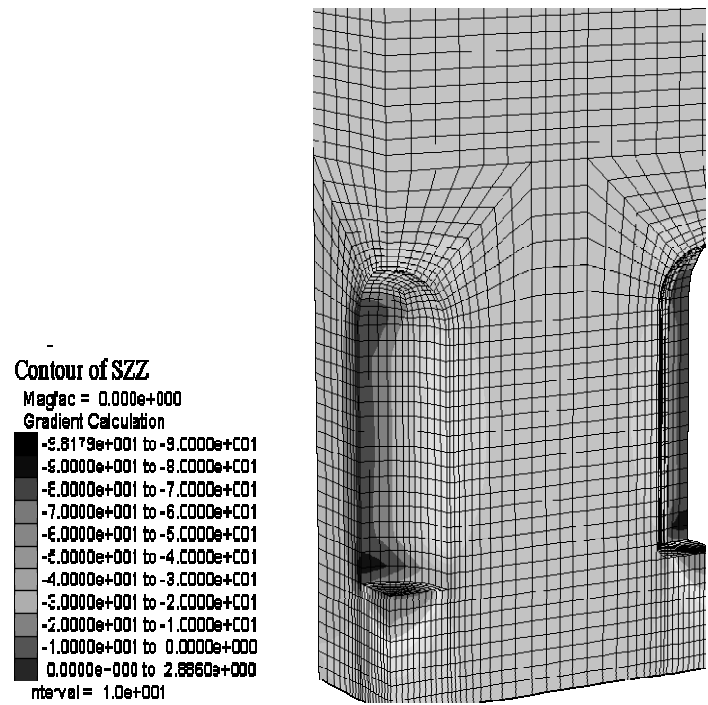

Fig7: Plot of contour of vertical stress for caverns in with cavern pressure of 20 MPa after one year. Where CTCD is 2.5 time cavern diameters.

## 5. Discussion

For natural gas storage in Qom region six geomechanical simulations were performed. Cavern deformation and bedding plane slip for varying cavern geometry, interbed properties, and roof beam properties analyzed.

A limited number of simulations have been completed to investigate and illustrate the relative influence of cavern height to diameter ratio, thickness of layers and salt on cavern deformation and bedding plane slip. These initial simulations support the following observations:

1. Non-salt interbeds across the cavern height provide areas of lower horizontal stress and subsequent lower resistance to hydraulic fracturing.

2. Caverns with height to diameter ratio (H/D) less than one produce high risk for roof beam interface slip, therefore risk level for well casing damage and roof caving increased.

3. Caverns with height to diameter ratio more than one, including non-salt interbeds along the height, and provide high risk for interface slip along those beds.

4. Additional non-salt interbeds near the center of the cavern have no any significant influence on the bedding plane slip in the roof interbed or in other non-salt interbeds near the cavern center. They merely seem to provide locations for additional slip.

5. Increased non-salt interbed thickness leads to increased slip with the adjacent salt, and

6. Increased salt beam thickness reduces risk for beam interface slip

## 6. Conclusions

Designing caverns for natural gas storage should take into account the mechanical properties of natural salt layer in order to perform accurate numerical simulations.

Numerical models are developed to analyze and determine the interaction of multiple caverns. In particular, the influence of the CTCD of multiple caverns on displacement and damages, are investigated.

The interaction of multiple horizontal salt caverns is evaluated to determine the minimum safe distance without compromising safety issues.

The geometric dimensions of each cavern are equal to the baseline case of the single cavern model. However, the important value is CTCD, which initially is selected as 60m (corresponds to 2 cavern diameters) Subsequently, the distance is increased to 75m, and 90m which correspond to 2.5 and 3 times cavern diameters. We find that 75m and 90m of caverns distance, the caverns are not affected by each other.